\documentclass[12pt]{iopart}

\usepackage{iopams,color,ulem}
\usepackage{xcolor}
\usepackage{graphics}
\usepackage{graphicx}
\usepackage{epsfig}

\definecolor{pdcolor}{rgb}{1,0.5,0}

\definecolor{pdblue}{rgb}{0,0,1}

\definecolor{rkgreen}{rgb}{0,1,0}

\begin{document}
\title[Age representation of L\'evy Walks]{Age representation of L\'evy Walks: partial density  waves,
relaxation and first passage time statistics}
\author{Massimiliano Giona$^1$, Mirko D'Ovidio$^2$, Davide Cocco$^2$,
Andrea Cairoli$^3$ and Rainer Klages$^{4,5*}$}

\address{$^1$ DICMA,
Facolt\`{a} di Ingegneria, La Sapienza Universit\`{a} di Roma,
via Eudossiana 18, 00184, Roma, Italy \\
$^2$Dipartimento SBAI,  La Sapienza Universit\`{a} di Roma,
via Antonio Scarpa 16, 00161 Roma, Italy\\
$^3$Department of Bioengineering,
Imperial College London, London SW7 2AZ United Kingdom\\
$^4$School of Mathematical Sciences, Queen Mary University of London,
London E1 4NS, United Kingdom \\
$^5$Institut f\"ur Theoretische Physik, Technische Universit\"at Berlin, Hardenbergstra{\ss}e 36, 10623 Berlin, Germany\\
$^*$ corresponding author }
\ead{r.klages@qmul.ac.uk}
\vspace{10pt}
\begin{indented}
\item[]
\end{indented}

\begin{abstract}
  L\'evy walks (LWs) define a fundamental class of finite velocity
  stochastic processes that can be introduced as a special case of
  Continuous Time Random Walks. Alternatively, there is a hyperbolic
  representation of them in terms of partial probability density
  waves. Using the latter framework we explore the impact of aging on
  LWs, which can be viewed as a specific initial preparation of the
  particle ensemble with respect to an age distribution. We show that
  the hyperbolic age formulation is suitable for a simple integral
  representation in terms of linear Volterra equations for any initial
  preparation. On this basis relaxation properties,
  {i.e., the convergence towards equilibrium of a
    generic thermodynamic function dependent on the spatial particle
    distribution, and first passage time statistics in bounded domains
    are studied by connecting the latter problem with solute release
    kinetics.}  We find that even {normal diffusive
    LWs, where the long-term mean square displacement increases
    linearly with time}, may display anomalous relaxation properties
  such as stretched exponential decay. We then discuss the impact of
  aging on the first passage time statistics of LWs by developing the
  corresponding Volterra integral representation. As a further natural
  generalization the concept of LWs with wearing is introduced to
  account for mobility losses.
\end{abstract}

\section{Introduction}
\label{sec1}

Since the unveiling of the mutual relationships between random motion
on microscopic scales and thermodynamic irreversibility described by
the diffusion equation \cite{gen}, statistical physics and the theory
of irreversible processes have taken great advantage from the
formulation of simple models of stochastic motion.  These models have
been widely used to understand the complex phenomenologies occurring
in fluids, colloidal and condensed matter systems especially when the
molecular structure (e.g.\ in polymer physics) or disorder (defects in
crystalline structures or amorphous materials) are accounted
for\ \cite{Kleinert1989}.

A huge field of investigation involves particle motion on discrete
lattices (see e.g.\ \cite{lrw1}), where both space and time become
discretized. Here particle motion is described with respect to an
operational time (discrete clock) attaining integer values, and the
distance between neighbouring sites of the lattice is fixed. The
transposition of the lattice model to physical reality requires the
definition of a characteristic length $\delta$ (spacing between
nearest neighboring sites) and time $\tau$ (time interval associated
with the elementary movement of a single operational clock).  This
class of models is suitable for coarse-graining leading to a
continuous statistical space-time representation by considering the
so-called hydrodynamic limit \cite{hydrolim1,hydrolim2}{.} We let
$\delta, \tau \rightarrow 0$ while imposing a specific constraint on
the behavior of $\delta$ and $\tau$, which is usually expressed by the
scaling condition $\delta^\alpha/\tau=\mbox{constant}$, where $\alpha$
is an integer.  In this way, the usual diffusion equation is recovered
from symmetric random walks (setting $\alpha=2$). Alternative
approaches are described in \cite{giona_phys_scripta}.  Lattice models
are particularly suited for including the effect of particle
interactions, either in the form of exclusion principles or as
interparticle potentials \cite{interaction1,interaction2}.  Recently
they have also been used to study collective motion in active matter
systems \cite{Tailleur1,Tailleur2}.  In addition lattice models
provide a clear pathway to analyze the core of fundamental problems
involving the foundations of statistical physics.  As an example, the
Kac ring model permits to address in an elegant and rigorous way the
relation between microscopic time-reversible motion, macroscopic
irreversibility and the role of a statistical description of the
dynamics \cite{kacring1,kacring2}.

Another basic paradigm of random kinematics originated from the
seminal article by Montroll and Weiss \cite{montroll}, which
introduced the concept of the {\it Continuous Time Random Walk}
(CTRW).  In this model the evolution of the system is still
parametrized with respect to an integer-valued operational time
{n} (counting the number of transitions in the particle
motion) {while the particle position and the physical
  time associated to it attain any real value}. Here the length
$\ell_n$ traveled and the time $\tau_n$ spent at the $n$-th transition
are real random variables (in most cases independent of each other),
which are characterized by a prescribed joint distribution functions.
This simple model triggered a huge flow of investigations
\cite{ctrw1,ctrw2,ctrw3,ctrw4,ctrw5} focusing primarily onto cases
where the main property of Brownian motion, namely the linear
long-term scaling of the mean square displacement as a function of
time, is broken yielding so-called anomalous diffusion \cite{KRS08}.

If the length $\ell_n$ and the time interval $\tau_n$ at the $n$-th
transition are not independent but linked to each other by the
existence of a characteristic and constant velocity $b$, the relation
between these two quantities can still be of a probabilistic nature,
$\ell_n = b \, \alpha_n \, \tau_n$, where $\alpha_n$ is a random
variable attaining values $\pm 1$ determining the direction of
motion. The resulting CTRW is usually referred to as a L\'evy Walk
(LW) {\cite{levy1,SWK87,KBS87}; see \cite{lwreview} for
  a review containing further details. The relation of LWs to discrete
  jump processes, using different scenarios for jump time and jump
  steps, is discussed in \cite{lwreview,magd_jump}.} The sequence of
transitions relating $\ell_n$ and $\tau_n$ can be made fully
deterministic by rewriting the LW dynamics as $\ell_n = s_0 \, b \,
(-1)^n \, \tau_n$, where the initial direction of motion $s_0$ is a
random variable.  The kinematics of a particle performing a LW is thus
specified by the distribution function of the time intervals $\tau_n$
between two subsequent transitions, which are assumed to be
independent of each other.  LWs are particularly attractive due to the
natural constraint of possessing a bounded propagation velocity, which
determines the almost everywhere regularity of their trajectories.  By
modulating the statistics of $\tau_n$ it is possible to provide simple
examples of random motions violating the Einsteinian linear scaling of
the mean square displacement with time \cite{zumofen1993}.

In the last two decades LWs found many useful applications in physical
and biological systems, from quantum dot fluctuations \cite{SHB09} to
the kinematics of unicellular microorganisms and cells
\cite{bactmotion2,HaBa12} and animal foraging \cite{VLRS11}; see
\cite{lwreview} for further applications.  A central issue in the
theory of LWs is the formulation of models for their statistical
characterization, expressed in the form of evolution equations for
their representative density functions (thus corresponding to
generalized Fokker-Planck equations) especially for those cases where
a LW displays anomalous diffusive properties
\cite{lw_frac1,lw_frac2,lw_frac3,lw_frac4}.

For trajectories of CTRWs a natural parametrization is obtained
through subordination.  We introduce: (a) a discrete operational time
$n$, which corresponds to the jumps occurring during the walk, and (b)
a discrete Markovian stochastic process in the operational time
$Y(n)$, which specifies the particle position after each jump.  Within
this picture, the physical time is expressed as a function of the
operational time via the elapsed time process $T(n)=\sum_{j=1}^n
\tau_j$.  Introducing now the process $N(t)\equiv \max\{n \geq 0:
T(n)\leq t\}$, the position of the walker can be expressed as
$X(t)=Y(N(t))$.  A similar relation can be shown to hold in the
continuum limit.  This formula immediately suggests that statistical
models for this random walk process can be naturally obtained by
considering exclusively the probability density function $P(x,t)$ of
finding the particle coordinate $X(t)$ at time $t$ in the interval
$(x,x+dx)$.  In fact, according to the previous formula, $P$ can be
expressed as the convolution of the corresponding densities for the
processes $N$ and $Y$.  Since the concept of a LW originated as a
branching of CTRW theory, a similar approach, focused on deriving an
evolution equation for the position statistics $P$, was later also
applied to LWs. In fact, despite the spatio-temporal coupling
introduced by imposing bounded particle velocities, these processes
are still Markovian in the operational time $n$.  However, for
anomalous diffusive LWs this modelling approach generates
convolutional operators corresponding to fractional derivatives of the
density $P(x,t)$, where, differently from the fractional derivatives
typically appearing in the evolution equations of CTRWs, the
spatio-temporal coupling manifests itself as advective derivatives and
retardation of $P$. Therefore, in a continuous time setting the
coordinate LW position process $X(t)$ is no longer a Markov process,
because the condition of bounded velocity and a fortiori the local
regularity of the trajectories enforces to add the local direction of
motion to the state description of the system. Exactly in the same
way a lattice random walk is not Markovian if the lattice spacing
$\delta$ and the hopping time $\tau$ are assumed to be finite and the
trajectory of a particle is interpolated between two transitions in a
continuous way \cite{giona_phys_scripta}.

In 2016 Fedotov published a short paper \cite{fedotov1} showing that
the statistical properties of a LW on the one-dimensional line are
fully described by a system of hyperbolic first-order differential
equations involving a system of partial probability densities (two in
the simplest case) accounting for the local direction of motion and
parametrized with respect to the particle age, which is defined as the
time elapsed from the latest transition in the direction of motion.
Further elaborations of this idea can be found in
\cite{fed1,fed2,fed3}.  The importance of this theoretical approach,
other than its technical value, is conceptual, as it stimulates a
radical change of paradigm in the parametrization of the trajectory of
a LW with respect to time. In fact, differently from the commonly
accepted picture stemming from CTRW theory, where the primitive time
is the operational time $n$ and the physical time $t$ should be
recovered from it, the statistical approach due to Fedotov puts the
physical time $t$ as the primitive temporal parametrization and
derives the statistical description in the physical space time by
using different analytical techniques.  Remarkably, such a seemingly
simple change of perspective yields a manifold of implications, as it
connects the theory of LWs with the classical approaches developed to
characterize stochastic processes possessing finite propagation
velocity, which originated from the articles of Goldstein
\cite{goldstein} and Kac \cite{kac1,kac2} and led later on to the
concepts of Poisson-Kac \cite{weiss,bena} and Generalized Poisson-Kac
processes \cite{gpk1,gpk2,gpk3}.  However, this useful relation comes
at the price of a seemingly increased complexity with respect to the
existing statistical approaches based on the evolution equation for
the overall probability density function $P(x,t)$, because an extra
independent variable must be introduced (the age) to parametrize the
partial densities of LWs. {Recently Poisson-Kac type
  models for active and biological particle motion have been
  considered under the diction of run-and-tumble models
  \cite{runtumble1}, including the case where Wiener perturbations are
  superimposed onto Poissonian perturbations \cite{runtumble2}. This
  case was also considered in \cite{gpk3}.}

{The aim of this article is to analyze the hyperbolic
  formulation of LW processes and the role played by the transitional
  age in them as an additional internal parameter to be introduced in
  order to completeley specify the representation of the local state
  of a LW particle. We then explore the consequences of this framework
  for the formulation of statistical theories of LW dynamics.  Along
  these lines we introduce} the concept of ``initial preparation'' in
the hyperbolic setting, and we show that many macroscopic properties
(with the sole exception of the long-term scaling of the moments) are
significantly influenced by it.  Furthermore, we show that, owing to
the simple first-order hyperbolic structure of the balance equations
for the partial probability densities, the additional level of
complexity can be ``renormalized out'' from the model by defining the
system evolution in terms of a single function $h(x,t)$ (or of two
functions $h_\pm(x,t)$ in the more general case) of a spatial $x$ and
a temporal $t$ coordinate. Consequently no extra degree of complexity
is added, other than the intrinsic convolutional nature of the
resulting integral equation, which is the fingerprint of a LW process.
The analysis of the concept of LW preparation solves the issue of
completeness in the description of a LW process, indicating that any
coarse model based exclusively on the overall density $P(x,t)$
corresponds to a specific initial preparation of the system involving
symmetries and constraints on the initial distribution of ages and
velocity directions.
 
The article is organized as follows: Section~\ref{sec2} introduces the
hyperbolic representation of LW statistics in terms of partial density
waves parametrized with respect to the transitional age and the
direction of propagation. This directly relates LWs to other classes
of processes possessing finite propagation velocity, such as
Poisson-Kac and Generalized Poisson-Kac processes
\cite{gpk1,gpk2,gpk3}. In Section~\ref{sec2_1} we show that the
transitional age formulation naturally leads to the concept of age
preparation of a LW ensemble out of which the notion of aging,
introduced for CTRWs first and later extended to LWs
\cite{Bar03,aging1,aging2,FrBa13,MJJCB14}, follows.  As a further
generalization, the concept of a {\it wearing LW} is introduced in
which the mobility properties of the process decay, by wearing, as a
function of the number of transitions.  Section~\ref{sec3} provides a
simple analytical representation of the solutions of the hyperbolic
system of equations for the partial densities characterizing the
statistics of LWs by reducing the problem to the solution of a simple
Volterra convolutional integral equation.  This approach allows us to
investigate the relaxational properties of LW fluctuations,
{namely the convergence towards equilibrium of generic
  thermodynamic quantities associated with the spatial distribution of
  the LW particle ensemble} in {closed} bounded domains
{equipped with reflective boundary conditions}. The
latter are discussed in Section~\ref{sec4}.  There we show that even
LWs that diffuse normally, i.e., with a position mean square
displacement scaling {linearly} for long time, may
display anomalous relaxation properties, such as a
Kohlrausch-William-Watts stretched exponential decay
\cite{stre1,stre2}.  Section~\ref{sec4} analyzes the influence of
different ensemble preparations on the first passage time statistics
in {closed} bounded domains by connecting this problem
with the release of a solute from a polymeric matrix. Finally,
Section~\ref{sec5} considers the application of the classical method
of images to the first passage time statistics, the validity of which
has been questioned in \cite{images} {in the case of
  L\'evy flights and LWs}. It is shown that the failure of the method
of images for the estimate of the first passage time statistics is a
generic feature of all the processes possessing finite propagation
velocity owing to the particular boundary condition at the passage
point, {see eq. (\ref{eq52_1})}, that cannot be matched
by the propagation of an additional symmetric point source. This is
due to the intrinsic lack of spatial symmetries of the elements of the
associated Green function matrix.

\section{Representation and age of L\'evy Walks}
\label{sec2}

LWs represent a prototype of stochastic processes possessing finite
propagation velocity, which under certain conditions can display
anomalous diffusive behavior in terms of a long-term deviation of the
mean square displacement from a linear Einsteinian scaling with time.
Throughout this article we consider one-dimensional LWs. The extension
to higher dimensions of the theory developed is in many cases
straightforward, in other cases less simple. In any case,
one-dimesional problems are definitely the proper framework for
addressing some fundamental physical concepts associated with the
representation of LWs, as will be shown in this article.

In a CTRW description of a LW, indicating with $x_n \in {\mathbb R}$
the particle position after the $n$-th transition in the direction of
motion and assuming a constant velocity $b$, the equations of motion
are given by
\begin{equation}
%\left \{
%\begin{array}{l}
x_{n+1} = x_n+ s_0 \, b \, (-1)^n \,  \tau_n , \qquad
t_{n+1} = t_n + \tau_n
%\end{array}
%\right . 
.
\label{eq2_1}
\end{equation}
Here $s_0$ is a random variable attaining values $\pm 1$ with equal
probabilities $1/2$ and $\tau_n$ are the time intervals between
subsequent transitions in the direction of motion, which
correspond to independent random variables defined by the
same probability density function $T(\tau)$. The random
variables $s_0$ and $\{\tau_h\}_{h=0}^\infty$ are independent of each
other so that, for any functions $f$ and $g$, $\langle f(s_0) \,
g(\tau_h) \rangle = \langle f(s_0) \rangle \, \langle g(\tau_h)
\rangle$, $\langle f(\tau_h) \, g(\tau_k) \rangle = \langle f(\tau_h)
\rangle \, \langle g(\tau_k) \rangle$, $h,k=0,1,\dots$, $h\neq k$,
where $\langle \cdot \rangle$ indicates the average with respect to
the corresponding probability measure. Note that Eq.~(\ref{eq2_1})
defines a special case of a CTRW where the direction of the velocity
alternates periodically in time. As mentioned previously, in an
alternative description the velocity itself is can be a random
variable.

In Eq.~(\ref{eq2_1}) the integer $n$ corresponds to the operational
time counting the transitions that determine a switch in the velocity
direction.  With respect to $n$ a LW is a Markov process for which the
probability density function $P_n(x)$ can be evaluated by employing
its Markovian structure.  The original definition of LW processes as
coupled CTRWs motivates the widely adopted strategy of defining
their statistical properties in terms of an evolution equation for the
position probability density function $P(x,t)$.

Conversely, the analysis of a LW process becomes more subtle when the
physical time $t \in {\mathbb R}^+$ is considered as the primitive
time parametrization, and the LW is viewed as a continuous process in
the independent variable $t$.  The simplest and most natural way of
defining this continuation is to adopt a Wong-Zakai interpolation
\cite{wong_zakai1,wong_zakai2} between two subsequent space-time
points $(x_n,t_n)$ and $(x_{n+1},t_{n+1})$ as prescribed by
eq.~(\ref{eq2_1}), i.e., by assuming that the kinematics of the LW is
described by means of straight line trajectories,
\begin{equation}
x(t)= x_n+ \frac{(x_{n+1}-x_n)}{(t_{n+1}-t_n)} \, (t-t_n) \, \,
\qquad t \in (t_n,t_{n+1}) \quad .
\label{eq2_2}
\end{equation}
Although other discontinuous interpretations of the kinematics of LW
processes have been considered \cite{lwreview,magd}, essentially in
the light of mathematical completeness it is rather clear that
eq.~(\ref{eq2_2}) represents the simplest and physically most
reasonable interpretation of the continuation of a LW trajectory,
which can capture the basic physical requirement of possessing a
finite propagation velocity and continuous trajectories.  However, the
continuous representation~(\ref{eq2_2}) renders the position process
$X(t)$ no longer a Markov process, because in the time-continuous
statistical description of these trajectories the local information on
the direction of motion becomes essential \cite{giona_phys_scripta}.

Therefore the local state of a LW process at time $t$ is defined by
the vector-valued state variable $(X(t),S(t),\tau(t))$, where $X(t)$,
$S(t)$ and $\tau(t)$ are the stochastic processes corresponding to the
particle position, the direction of motion and the transitional age of
the particle, respectively.  The process $S$ attains values $\pm 1$,
depending on whether the particle is moving towards positive
$x$-values ($S(t)=+1$) or negative ones ($S(t)=-1$).  The transitional
age is defined as the time elapsed from the latest transition in the
velocity direction.  Remarkably, this formalism allows to consider
generic statistics for the transition times, which we call $T(\tau)$,
and not only purely exponential distributions.

By considering the triplet $(x,s,\tau)$ a LW process is brought back
to the Markovian realm. Indeed, its conditional probability density
function $p(x,s,\tau,t \, | \, x_0,s_0,\tau_0,t_0)$, with $t>t_0$,
satisfies a Chapman-Kolmogorov equation out of which the corresponding
Fokker-Planck equation for its probability density function
$p(x,s,\tau,t)$ can be derived. Since the velocity direction $s$
attains the values $\pm 1$, $p(x,s,\tau,t)$ can be split into two
partial densities $p_\pm(x,t;\tau)=p(x,\pm1,\tau,t)$, which
corresponds to the system of partial probability density functions
adopted by Fedotov \cite{fedotov1} in order to describe, in the most
general way, the statistical evolution of one-dimensional LWs. The
application of the Chapman-Kolmogorov equation in this case leads to a
system of hyperbolic evolution equations for the partial probability
densities \cite{fedotov1,fed1}
\begin{equation}
\frac{\partial p_\pm(x,t;\tau)}{\partial t}=
\mp b \, \frac{\partial p_\pm(x,t;\tau)}{\partial x} - \frac{\partial p_\pm(x,t;\tau)}{\partial \tau} - \lambda(\tau) \, p_\pm(x,t;\tau) \quad ,
\label{eq2_3}
\end{equation}
where $\lambda(\tau)$ is the transition rate at age $\tau$, i.e.,
$\lambda(\tau) \, d t$ is the probability that a LW particle with age
$\tau$ will perform a switching in the direction of motion in
the time interval $(t,t+dt)$.  The transition rate $\lambda(\tau)$ is
related to the transition time probability density $T(\tau)$ by
\begin{equation}
T(\tau)= \lambda(\tau) \, e^{-\Lambda(\tau)} \quad ,
\quad \Lambda(\tau)=\int_0^\tau \lambda(\tau^\prime) \, d \tau^\prime\quad .
\label{eq2_4}
\end{equation}
The effect of the transitions regarding the parametrization with
respect to $\tau$ of the particle ensemble are accounted for by the
boundary condition at $\tau=0$. This is formulated such that all the
particles, which at any time $t$ and position $x$ performed a
transition in the direction of the velocity, return to a vanishingly
small transitional age with reversed velocity direction, i.e.,
\begin{equation}
p_\pm(x,t;0)= \int_0^\infty \lambda(\tau) \, p_\mp(x,t;\tau) \, d \tau . 
\label{eq2_5}
\end{equation}
Eqs. (\ref{eq2_4})-(\ref{eq2_5}) represent the partial density
approach to the statistical characterization of LWs first developed in
\cite{fedotov1}. We show below that this formalism paves the way to a
significant improvement in the understanding of the properties of LWs
by motivating a shift of paradigm with surprising consequences on the
theory of LWs.

At first sight, it may appear that this description is significantly
more complex than the coarse approach based
exclusively on the overall density function
\begin{equation}
P(x,t)= \sum_{\alpha=\pm} \int_0^\infty p_\alpha(x,t;\tau) \, d \tau
\label{eq2_6}
\end{equation}
that can be expressed in terms of integer or fractional-order
operators with retardation effects in $P$
\cite{lw_frac1,lw_frac2,lw_frac3,lw_frac4}.  As a point of fact the
partial density approach and the coarse description serve two
different purposes. The pair $(s,\tau)$ as internal variables of a LW
process provides a complete description of its internal degrees of
freedom. This brings back the concept of ``{\it preparation}'', which
is further discussed below. Conversely, a coarse model for the overall
density $P(x,t)$ should be viewed as a long-term model accounting for
the qualitative scaling properties of the dynamics, once the internal
dynamics of a LW involving the redistribution amongst the two velocity
directions and amongst the transitional ages has reached an
equilibrium condition (at least in those cases where an equilibrium
exists). However, as shown in the next section, even in the case of
the partial wave formulation it is possible to derive a simple linear
integral equation of convolutional type involving solely an auxiliary
function $h(x,t)$ of two variables, exactly as for the overall coarse
model, out of which all the properties regarding the space-time
evolution of the LW can be obtained.

There is another major merit of the partial density formulation, as it
provides a formal unification of several classes of stochastic
processes possessing finite propagation velocity within a unique
hyperbolic description of their statistical properties. This is the
case of Poisson-Kac (PK) processes for which $\lambda(\tau)=\lambda=
\mbox{constant}$, thus implying a Markovian transition amongst the
ages described by an exponential density function $T(\tau)= \lambda \,
e^{-\lambda \, \tau}$.  From eqs.~(\ref{eq2_3}) and (\ref{eq2_5}) and
by setting $\widehat{P}_\pm(x,t)= \int_0^\infty p_\pm(x,t,\tau) \, d
\tau$, the partial densities $\widehat{P}_\pm(x,t)$, uniquely
parametrized with respect to the local velocity direction, satisfy the
equations
\begin{equation}
\frac{\partial \widehat{P}_\pm(x,t)}{\partial  t}= \mp b\frac{\partial \widehat{P}_\pm(x,t)}{\partial x} \mp\lambda \, \left [ \widehat{P}_+(x,t)-\widehat{P}_-(x,t)
\right ]
\label{eq2_7}
\end{equation}
out of which a single equation for the overall density can be derived
if required (in this case of Cattaneo type \cite{kac2}). In the case of
PK processes the transitional age formalism can be defined, but
information on the age distribution is completely irrelevant in the
statistical evolution of the process, as the only internal parameter
that counts is the local direction of motion $s(t)=\pm1$.

This formal equivalence is however extremely useful, as concepts and
methods developed for PK processes can be fruitfully transferred to
the analysis of LWs. For example, it has been shown in
\cite{giona_pucci} that problems arise when studying PK processes in
bounded domains (e.g.\ an interval in one dimension), associated with
the proper setting of the boundary condition in order to fulfil the
requirement of positivity of the resulting probability distributions.
This is also the case of the maximum flux condition found in
\cite{flux1,flux2}, which yields a straightforward explanation within
the partial density representation \cite{brasiello}. We will exploit
this analogy in Sec.~\ref{sec5} when discussing the first passage time
problem for LWs.

\section{Preparation and aging of L\'evy Walks}
\label{sec2_1}

Consider again the CTRW description of a LW at discrete time instants
corresponding to the points in time at which transitions in the
direction of motion occur. In this framework it is implicitly assumed
that the initial time $t=0$ yields the instant at which all the
particles have just performed a transition and that the initial
directions of motion are distributed amongst $s=\pm1$ in an
equiprobable way. Viewed with respect to the partial density formalism
this represents indeed a very peculiar case of initial condition.  As
discussed above, the couple $(s,\tau)$ describes the internal degrees
of freedom of a LW process, and consequently the initial state of the
system should be defined also with respect to these variables in order
to completely characterize the process.

This observation leads to the concept of {\it preparation} of an
ensemble of LW particles/fluctuations, which can be viewed as the
specification of the initial state of the ensemble with respect to the
internal parameters $(s,\tau)$.  In symbols, the preparation of a LW
process is just the pair
$(\{\pi_\alpha^0,\phi_\alpha^0(\tau)\}_{\alpha=\pm})$.  On the one
hand, ${\pi_\alpha^0}$ are the probabilities associated with the
initial distribution of the directions of motion, which thus satisfies
the evident properties $\pi_\alpha^0 \geq 0$, $\pi_+^0+\pi_-^0=1$.  On
the other hand, $\phi_\alpha^0(\tau)$ are density functions accounting
for the distribution with respect to the transitional age of the two
subpopulations of particles, which thus satisfies $\phi_\alpha(\tau)
\geq 0$, $\int_0^\infty \phi_\alpha(\tau) \, d \tau=1$, $\alpha=\pm$.
Consequently, assuming that all the particles are initially located at
$x=0$, the initial condition specifying the solutions of the
hyperbolic eqs.~(\ref{eq2_3}) and (\ref{eq2_5}) takes the form
\begin{equation}
p_\pm(x,0;\tau)=p^0_\pm(x,\tau) = \pi_\pm^0 \, \phi_\pm^0(\tau)\,
\delta(x) \quad .
\label{eq2_8}
\end{equation}
{The hyperbolic formulation of LW dynamics
permits to consider generic expressions for the  initial
age-distribution of a LW ensemble.}
For instance, the CTRW preparation of a LW ensemble is just
$\pi_\pm^0=1/2$, $\phi_\pm^0(\tau)=\delta(\tau)$ corresponding to an
equiprobable and impulsive initial distribution with all the particles
possessing vanishing transitional age.

In this perspective the concept of aging introduced for CTRWs and
extended to LWs \cite{Bar03,aging1,aging2,FrBa13,MJJCB14} follows
naturally as a particular preparation of the system. An aged LW system
possessing aging time $t_a$ is an ensemble of LW particles that has
evolved for a time interval $t_a$ starting from the CTRW
preparation. Because the dynamical evolution of the particle ensemble
under consideration preserves symmetries the directions of motion are
equiprobable, the age densities $\phi_\pm(\tau)$ coincide with each
other, and they are equal to
\begin{equation}
\phi_\pm(\tau)= \widehat{\pi}(t_a,\tau)\quad ,
\label{eq2_9}
\end{equation}
where $\widehat{\pi}(t,\tau)\equiv \int_{-\infty}^{\infty}
p_{\pm}(x,t;\tau)\mathrm{d}x$ is the solution of the age-dynamics
\begin{equation}
\frac{\partial \widehat{\pi}(t,\tau)}{\partial t}= - \frac{\partial \widehat{\pi}(t,\tau)}{\partial \tau} -\lambda(\tau) \, \widehat{\pi}(t,\tau)
\label{eq2_10}
\end{equation}
equipped with the boundary and initial conditions
\begin{equation}
\widehat{\pi}(t,0)=\int_0^\infty \lambda(\tau) \, \widehat{\pi}(t,\tau) \, d \tau\quad , \quad
\widehat{\pi}(0,\tau)=\delta(\tau) \quad .
\end{equation}

However, this concept of preparation is broader than aging. It is
intrinsically associated with the age representation of LWs and finds
an immediate interpretation in those cases where a LW represents a
model for complex fluctuations in condensed and soft matter
physics. Examples are glasses or polymeric solutions, where memory
effects strongly influence the transport and relaxation properties of
the system.  In this case the two parameters velocity $b(\Theta)$ and
transition rate $\lambda(\tau,\Theta)$ characterizing a LW can depend
on the physical temperature $\Theta$. Assume for simplicity that the
LW is transitionally ergodic in the range of temperatures considered,
which means that the age dynamics (\ref{eq2_10}) admits, for constant
$\Theta_0$, an invariant density, $\widehat{\pi}^*(\tau;\Theta_0)= A
\, e^{-\Lambda(\tau;\Theta_0)}$, where $\Lambda(\tau;\Theta_0)$ is
defined by eq.~(\ref{eq2_4}) with $\lambda(\tau)$ replaced by
$\lambda(\tau;\Theta_0)$, and $A$ is a normalization constant.  Next,
suppose that at $t=0^+$ the system temperature is changed abruptly,
setting it to $\Theta\neq \Theta_0$. In this scenario the dynamics of
the system at temperature $\Theta$ is significantly influenced by the
previous preparation, which corresponds to the equilibrium conditions
at the initial temperature $\Theta_0$.
 
The latter observation suggests a further generalization of LW
dynamics. The preparation and the aging effects are a manifestation of
the initial condition on the structure of the internal degrees of
freedom that influence the short to intermediate time scales or the
properties in bounded systems (see Sec.~\ref{sec4}), as it impacts on
the complex transition mechanism of LWs, which in general is
characterized by long-range memory effects. Another generalization
borrowed from condensed matter physics and material science may
involve the fact that LW fluctuations in complex materials (glasses)
may be subjected to a progressive {\it wearing} as a function of time
that modifies the mobility properties, resulting in a progressive
decrease of the effective velocity $b$.  Therefore, we define a {\it
  Wearing L\'evy Walk} (WLW) as a LW whose trajectories are given by
eqs.~(\ref{eq2_1}) and (\ref{eq2_2}) for which the velocity $b$ is no
longer constant but depends on the number of transitional events
experienced, i.e., on the operational time $n$ entering
eq.~(\ref{eq2_1}), i.e.,
\begin{equation}
b=b_0 \beta(n) \, , \qquad \beta(n+1) \leq \beta(n), \, \qquad n=0,1,\dots \quad .
\label{eq2_11}
\end{equation}
{Related models have been discussed in
  \cite{SWK87,KBS87} and very recently in \cite{AlRa18,BSS19}.} If the
wearing process is sufficiently slow, it may occur that the system
still maintains some level of fluctuation even in the long run,
characterized by qualitatively different properties with respect to
the case where the wearing dynamics is absent. In order to give an
example of this phenomenon let $b_0=1$ [a.u.] and {the
  transition rate $\lambda(\tau)$ defined in eq.~(\ref{eq2_3})}
\begin{equation}
\lambda(\tau)= \frac{\xi}{1+\tau} \; \; [a.u.] \quad ,
\label{eq2_12}
\end{equation}
{which according to eq.~(\ref{eq2_4}) yields the
  transition time probability density $T(\tau)=\xi/(1+\tau)^{\xi+1}$.}
For $\xi\leq 1$ the system is not transitionally ergodic,
{since no equilibrium age distribution exists.
  Indicating with $R^2(t)$ the mean square displacement at time $t$,
  $R^2(t)=\langle x^2(t)\rangle-\langle x(t) \rangle^2$, $R^2(t) \sim
  t^2$. For $\xi>1$ the LW is transitionally ergodic, and the
  equilibrium age density exists and is given by $\widehat{\pi}_{\rm
    equil}(\tau)= A \, e^{-\Lambda(\tau)}$, where $A$
  is a normalization constant.  For $1<\xi<2$ it is characterized} by
anomalous diffusive behavior providing a superdiffusive scaling of
$R^2(t) \sim t^{3-\xi}$ while $R^2(t) \sim t$ for $\xi>2$
\cite{fedotov1}. If a slow wearing mechanism is added, by assuming for
the function $\beta(n)$ a logarithmic behavior
\begin{equation}
\beta(n)=\frac{1}{1+\log(1+n)}
\label{eq2_13}
\end{equation}
the transport properties change in a qualitative way.
\begin{figure}[h]
\begin{center}
{\includegraphics[height=6.5cm]{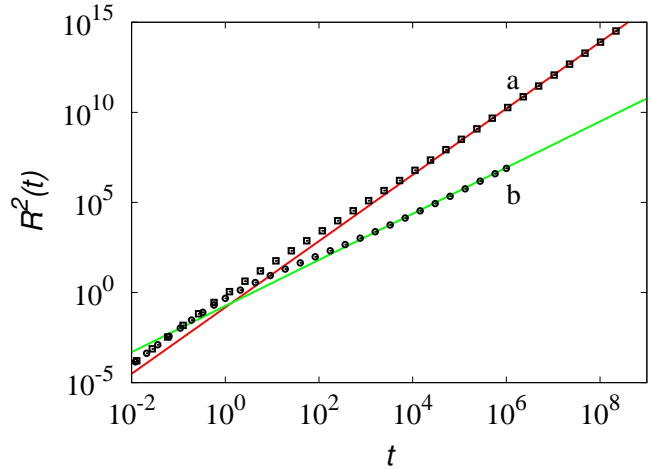}}
\end{center}
\caption{The mean square displacement $R^2(t)$ vs.\ time $t$ for the
  Wearing LW defined by eqs.~(\ref{eq2_11})-(\ref{eq2_13}). Dots are
  the results of stochastic simulations, lines represent {the  LW
  long-term scaling} $R^2(t)\sim t^\gamma$. Line (a) and ($\square$)
  refer to $\xi=0.5$ for which $\gamma=1.84$, line (b) and ($\circ$)
  to $\xi=1.5$ for which $\gamma=1.28$.}
\label{Fig1}
\end{figure}
Figure \ref{Fig1} depicts the scaling of $R^2(t)$ as a function of
time $t$ for the WLW at the two different values of
$\xi=0.5,\,1.5$. Simulations have been performed using an ensemble of
$10^7$ particles. While for small time we observe the expected
ballistic scaling, in both cases a long-term power-law scaling is
observed, $R^2(t) \sim t^\gamma$, with an exponent different from the
case without wearing: $\gamma=1.84 \pm 0.02$ at $\xi=0.5$, whereas
instead the classical LW would predict $\gamma=2$, and $\gamma=1.28
\pm 0.03$ at $\xi=1.5$, with $\gamma=1.5$ instead in the absence of
wearing.

A more detailed analysis of WLWs falls outside the scope of this
article and will be addressed in forthcoming works. What is important
to notice here is that once the age structure and formalism of LWs is
assumed, {generalizations and extensions of the
  internal age parametrization follow systematically and} can be
exploited for adapting the LW paradigm to the complexity of physical
phenomenology.

\section{Integral representation of the solutions}
\label{sec3}

Here we show that the partial density formalism expressed by
eqs.~(\ref{eq2_3}) and (\ref{eq2_5}) provides the same level of
analytical complexity than any other model based on the formulation of
an evolution equation for the overall probability density
$P(x,t)$. The approach followed is similar to a corresponding analysis
developed in \cite{fedotov1}, where a single evolution equation for
$P(x,t)$ was finally obtained by enforcing an initial preparation of
CTRW-type.  Below, starting from the hyperbolic formulation for any
initial preparation, we derive a single integral equation for an
auxiliary function, which depends solely on a spatial and temporal
variable.

Consider the propagation of a LW on the real line, defined
statistically by eqs.~(\ref{eq2_3}) and (\ref{eq2_5}) and equipped
with the initial conditions $p_\pm(x,0;\tau)=p^0_\pm(x,\tau)$.  Assume
the following initial symmetry
\begin{equation}
p_+^0(x,\tau)=p_-^0(-x,\tau) \quad \forall x \in {\mathbb R}
 \,, \; \tau \geq 0\quad ,
\label{eqx4}
\end{equation}
which is the symmetry characterizing the CTRW preparation or the
initial setting of a LW with aging (see Sec.~\ref{sec2_1}). This
symmetry involves solely the initial distribution of velocity
directions and not the initial age distribution, which remains
generic.

In this case, due to the symmetric propagation towards positive/negative
$x$ values of the forward ($p_+(x,t;\tau)$) and backward
($p_-(x,t;\tau)$) densities, one has
\begin{equation}
p_-(x,t;\tau) = p_+(-x,t,\tau)\quad .
\label{eqx5}
\end{equation}
Consequently, in the analysis of the process it is sufficient to
consider solely the forward density $p(x,t;\tau)=p_+(x,y;\tau)$, whose
evolution equation becomes nonlocal in space according to
\begin{eqnarray}
\frac{\partial p(x,t;\tau)}{\partial t} & = & - b \, \frac{\partial p(x,t;\tau)}
{\partial x} - \frac{\partial p(x,t;\tau)}{\partial  \tau} \nonumber 
-\lambda(\tau) \, p(x,t;\tau)\\
p(x,t;0)& = & \int_0^\infty \lambda(\tau) \, p(-x,t;\tau) \, d \tau
\label{eqx6} \\
 p(x,0,\tau) &= & p^0(x,\tau)=p_+^0(x,\tau) \nonumber \quad .
\end{eqnarray}
Observe that the nonlocality in eq.~(\ref{eqx6}), is not a physical
property but rather a mathematical superstructure introduced in order
to enforce the symmetries and to get rid of the backward density wave.
Obviously, the overall density $P(x,t)$ is given by
\begin{equation}
P(x,t)= \int_0^\infty \left [ p(x,t;\tau)+p(-x,t;\tau) \right ] \, d \tau \quad .
\label{eqx7}
\end{equation}
Consider then the transformation
\begin{equation}
p(x,t;\tau) = e^{-\Lambda(\tau)} \, q(x,t;\tau) \quad .
\label{eqx8}
\end{equation}
From eq.~(\ref{eqx6}), the  equation for $q(x,t;\tau)$ becomes
\begin{equation}
\frac{\partial q(x,t;\tau)}{\partial t}= - b \, \frac{\partial q(x,t;\tau)}{\partial x} - \frac{\partial q(x,t;\tau)}{\partial \tau}
\label{eqx10}
\end{equation}
equipped with the boundary and initial conditions
{
\begin{eqnarray}
q(x,t;0) &= & \int_0^\infty T(\tau) \, q(-x,t,\tau) \, d \tau \nonumber \quad ,\\
 q(x,0;\tau) &= & q^0(x,\tau)= e^{\Lambda(\tau)} \, p^0(x,\tau)\quad ,
\label{eqx11}
\end{eqnarray}
}
{where $T(\tau)$  and $\Lambda(\tau)$ are defined by eq.~(\ref{eq2_4}).}
Equation~(\ref{eqx10}) is a first-order constant coefficient equation
casted in a conservation form that can be solved with the method of
characterics: Its solution attains the form
\begin{equation}
q(x,t;\tau) = \phi(x-b \, t,t-\tau)\quad .
\label{eqx13}
\end{equation}
By considering the boundary condition at $\tau=0$, it follows that for
$\tau \geq t$ $q(x,t;\tau)$ consists solely of the propagation of the
initial condition both in space and age.  Conversely, for $\tau<t$ the
solution can be formally expressed by introducing an auxiliary
function $h(x,t)$.  Thus, eq.~(\ref{eqx13}) can be written as
\begin{equation}
q(x,t;\tau)=
\left \{
\begin{array}{lll}
q^0(x-b\, t, \tau-t) & & \tau \geq t \\
h(x-b \, \tau, t-\tau) & & \tau <t
\end{array}
\right .
\label{eqx14}
\end{equation}
Substituting eq.~(\ref{eqx14}) into the boundary condition
(\ref{eqx11}), the equation for $h(x,t)$ follows
\begin{eqnarray}
h(x,t) &=& \int_0^t T(\tau) \, h(-x-b \, \tau,t-\tau)
\, d \tau \nonumber \\
&+& \int_t^\infty T(\tau) \, q^0(-x-b \, t,\tau-t)
\, d \tau\quad ,
\label{eqx15}
\end{eqnarray}
{which holds for $t>0$.} The latter equation can be obtained in terms of the initial
condition $p^0(x,\tau)$ to
\begin{eqnarray}
h(x,t) &  = & \int_0^t T(\tau) \, h(-x-b \, \tau,t-\tau)
\, d \tau    \nonumber \\
& + &\int_t^\infty T(\tau) \,  e^{\Lambda(\tau-t)} \, p^0(-x-b \, t,\tau-t)
\, d \tau
\label{eqx16}
\end{eqnarray}
 and the density $p(x,t;\tau)$ is thus given by
\begin{equation}
p(x,t;\tau) =
\left \{
\begin{array}{lll}
e^{-\Lambda(\tau)} \, e^{\Lambda(\tau-t)} \, p^0(x-b \, t,\tau-t) & & \tau \geq t \\
e^{-\Lambda(\tau)} \, h(x-b \, \tau,t-\tau) & & \tau< t\quad .
\end{array}
\right .
\label{eqx17}
\end{equation}
{Since $h(x-b \, \tau,t-\tau)$ is defined stictly for $t>\tau$,
it can always be set to $h(x,t)=0$ for any $x$, $t\leq 0$. 
Eq.~(\ref{eqx16}) can be expressed equivalently as}
\begin{equation}
h(x,t)= \int_0^t T(t-\tau) \, h(-x-b \, t+ b \, \tau,\tau) \, d \tau \, ,
+ G^0(t)
\label{eqx19}
\end{equation}
where the forcing term $G^0(t)$ is a linear functional of the initial
condition $p^0(x,\tau)$ corresponding to the second integral on the
r.h.s.\ of eq.~(\ref{eqx16}).  If the symmetry condition (\ref{eqx5})
is removed it is still possible to derive an integral representation
of the solutions involving two auxiliary functions $h_\pm(x,t)$.  This
is addressed in Sec.~\ref{sec5} in connection with the analysis of the
first passage time problem. Several observations follow from {the above}
derivation:
\begin{itemize}
\item By applying the method of characteristics it is possible to
  compress all the physical information about the spatial-temporal
  propagation of a LW into a single function $h(x,t)$ of the two
  arguments $x \in {\mathbb R}$ and $t\geq 0$, analogously to the
  evolution equation associated with the overall density function
  $P(x,t)$ involving, for some $\lambda(\tau)$, fractional
  derivatives.
\item Eq.~(\ref{eqx16}) is exact and holds for any initial
  preparation of the system and any functional form of
  $\lambda(\tau)$.
\item The memory effects of the age dynamics characterizing a LW can
  be clearly appreciated by the convolutional nature of the first term
  entering eq.~(\ref{eqx19}), which is a linear Volterra integral
  equation whose kernel is the transition time density $T(\tau)$. Due
  to the simultaneous propagation both in space and along the ages,
  this convolution involves both arguments of the function $h(x,t)$.
\end{itemize}

\begin{figure}[h]
\begin{center}
{\includegraphics[height=12.5cm]{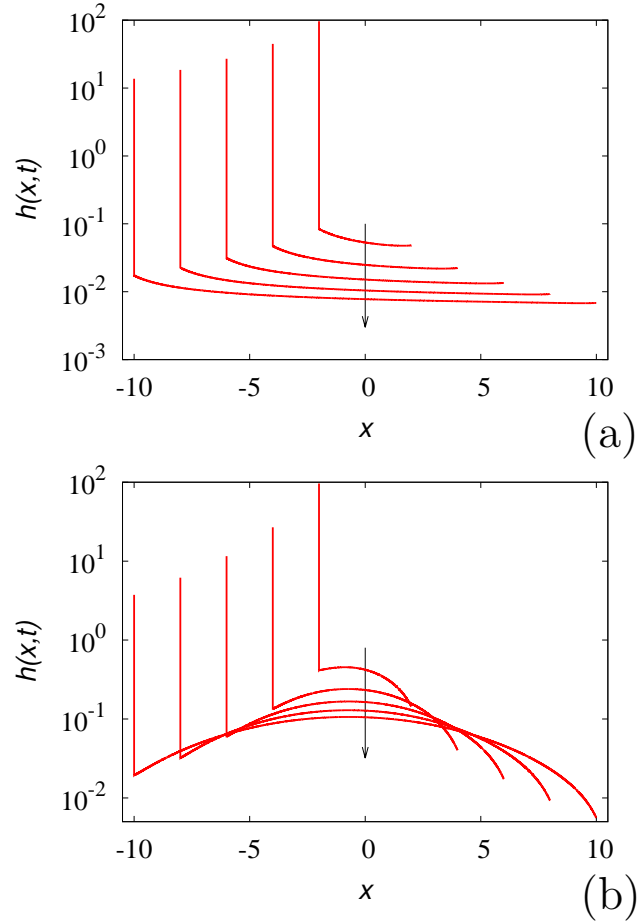}}
\end{center}
\caption{Function $h(x,t)$ vs.\ $x$ obtained from the numerical
  solution of the Volterra integral equation for the LW defined by the
  transition rate eq.~(\ref{eq2_12}). Panel (a) refers to $\xi=0.5$,
  panel (b) to $\xi=1.5$. The arrows indicate increasing time instants
  $t=2,\,4,\,6,\,8,\,10$.}
\label{Fig2}
\end{figure}

\begin{figure}[h]
\begin{center}
{\includegraphics[height=12.5cm]{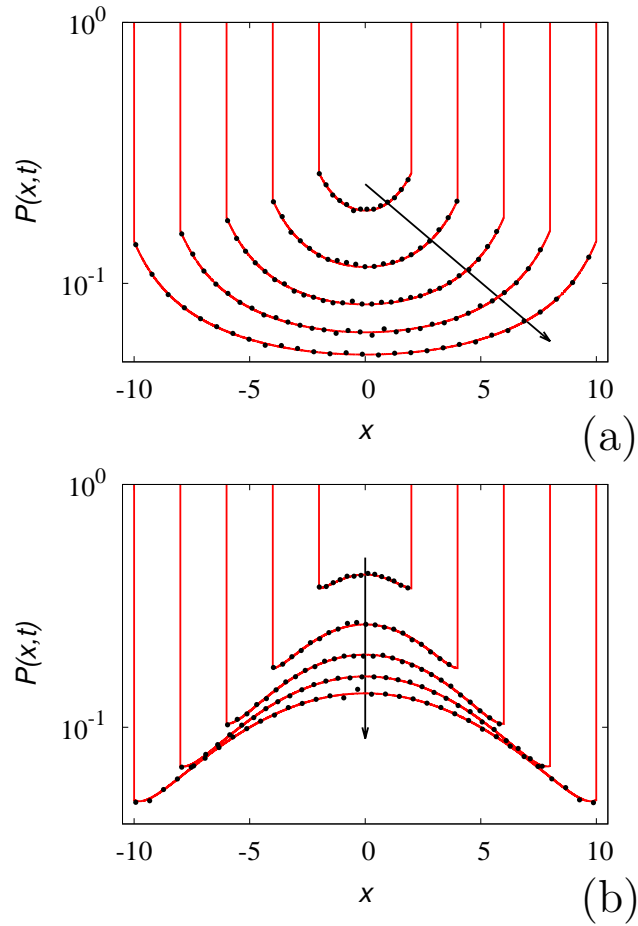}}
\end{center}
\caption{Overall density $P(x,t)$ vs.\ $x$ derived for the
  $h$-function depicted in Fig.~\ref{Fig2}.  Panel (a) refers to
  $\xi=0.5$, panel (b) to $\xi=1.5$. The arrows indicate increasing
  time instants $t=2,\,4,\,6,\,8,\,10$. Symbols represent the results
  of corresponding stochastic simulations of the process.}
\label{Fig3}
\end{figure}

The representation (\ref{eqx16}) is amenable to a simple numerical
integration. For simplicity set $b=1$ [a.u.] and use an impulsive
initial condition, $p^0(x,\tau)=\delta(x) \, \delta(\tau)$.  For this
particular case, the prefactor of $p^0(x-b \, t,\tau-t)$ in
eq. (\ref{eqx17}) simplifies as $e^{-\Lambda(t)}$. Assuming equal step
size for $x$, $t$ and $\tau$, i.e., $\Delta x= \Delta t= \Delta \tau$,
and defining the grid approximation $\widehat{h}[m,n]=h(x_m,t_n)$,
$m=\dots,-1,0,1,\dots$, $n=0,1,\dots$, where $x_m= m \, \Delta x$,
$t_n=n \, \Delta t$, the simplest discretization of eq.~(\ref{eqx19})
provides the solution algorithm
\begin{equation}
\widehat{h}[m,n]= \sum_{i=0}^{n-1} \widehat{T}[n-i]  \, \widehat{h}[-m-n+i,i] \, \Delta \tau+
\widehat{T}[n] \, \widehat{\delta}[m+n]\quad ,
\label{eqx20}
\end{equation}
where $\widehat{T}[k]=T(\tau_k)$, $\tau_k=k \, \Delta \tau$, and
$\widehat{\delta}[h+k]$ is the numerical approximation for a Dirac
delta function, $\widehat{\delta}[k]=0$ for $k \neq 0$,
$\delta[0]=1/\Delta x$. For any $n$, $\widehat{h}[m,n]$ is different
from zero solely for $|m| \leq n$.  The overall density function
$\widehat{P}[m,n]=P(x_m,t_n)$ at time instant $t_n$ is thus written
as $\widehat{P}[m,n]=\widehat{\pi}[m,n]+\widehat{\pi}[-m,n]$, where
\begin{equation}
\widehat{\pi}[m,n]=\sum_{i=0}^{n-1} \widehat{\Lambda}[i] \, \widehat{h}[m-i,n-i] \, \Delta \tau + \widehat{\Lambda}[n] \,  \widehat{\delta}[m-n]
\label{eqx21}
\end{equation}
and $\widehat{\Lambda}[k]=exp(-\Lambda(\tau_k))$. To give a numerical
example, Fig.~\ref{Fig2} depicts the evolution of the $h$-function of
the LW model defined by eq.~(\ref{eq2_12}) at two different values of
the parameter $\xi$ by applying eq.~(\ref{eqx20}) with $\Delta
x=10^{-3}$. The corresponding overall density profiles $P(x,t)$,
derived from the $h$-function via eq.~(\ref{eqx21}) and normalized to
unity, are depicted in Fig.~\ref{Fig3} by comparing them with
stochastic simulations of the corresponding problem, obtained using an
ensemble of $N_p=10^8$ particles.  {The markedly
  different behaviour for these two values of $\xi$ corresponds to the
  fact that for $\xi<1$ the LW is not transitionally ergodic i.e., no
  transitional age equilibrium distribution exists, while it does for
  $\xi>1$, where the transitional age equilibrium distribution is
  given by $\widehat{\pi}_{\rm equil}(\tau)=A \, e^{-\Lambda(\tau)}$
  with normalization constant $A$. In the present case
  $\widehat{\pi}_{\rm equil}(\tau)= (\xi-1)/(1+\tau)^\xi$, $\xi>1$.
  This manifests itself in the different convexity of the distribution
  between the ballistic peaks, see Ref.~\cite{lwreview} for plots of
  these different distributions.}

\section{Problems in bounded domains: relaxation and diffusional release
 dynamics}
\label{sec4}

The influence of the internal preparation of a LW ensemble controls
the short to intermediate scale properties and the statistical
behavior in bounded systems. Let us address these issues with some
examples. 
{\subsection{Relaxational dynamics}}
\label{sec4_1}

Consider the evolution of LW fluctuations in a bounded
closed domain, which in the one-dimensional case can be represented by
the interval $[0,L]$. The system of hyperbolic equations
(\ref{eq2_3}), defined for $x \in (0,L)$, is thus equipped with
reflective boundary conditions at the endpoints
\begin{eqnarray}
p_+(0,t;\tau) & = & p_-(0,t;\tau) \nonumber \\
p_-(L,t;\tau) & = & p_+(L,t;\tau)
\label{eq4_1}
\end{eqnarray}
for any $t>0$, $\tau>0$ corresponding to the total reflection of the
incoming wave at the boundaries where it inverts its direction of
propagation: at $x=0$ the incoming wave is $p_-(0,t;x)$, at $x=L$ it
is $p_+(L,t;\tau)$. Independently of the transitionally ergodic nature
of the age dynamics, the spatial distribution becomes asymptotically
uniform, i.e., the overall density function $P(x,t)$ approaches the
uniform density $P^*(x)=1/L$, $x \in (0,L)$ for $x \in (0,L)$
corresponding to the equilibrium distribution, at least restricted to
the spatial dynamics.

Let $f(x)$ be any thermodynamic function associated with the LW
fluctuations and $\overline{f}^*$ its equilibrium value with respect
to the long-term limit density $P^*(x)$, i.e., $\overline{f}^*=
\int_0^L f(x) \, P^*(x) \, d x$.  The relaxation function $R_f(t)$
referred to the observable $f(x)$ is therefore the absolute value of
the difference of the average value of $f(x)$ at time $t$ and its
(long-term) equilibrium value $\overline{f}^*$,
\begin{equation}
R_f(t)= | \langle f(x)\rangle(t) - \overline{f}^* |
= \left | \int_0^L f(x) \, P(x,t) \, dx - \frac{1}{L}
\int_0^L f(x) \, d x \right | \quad .
\label{eq4_2}
\end{equation}
The reflective conditions (\ref{eq4_1}) do not alter the age structure
of the LW process so that the analysis developed in the previous
section, at least regarding the age dynamics, can be qualitatively
applied to the present case.  Suppose that the LW system is prepared
in a CTRW-way with an impulsive initial age distribution
$\delta(\tau)$. Consequently, during the relaxation process of the
spatial distribution $P(x,t)$ towards $P^*(x)$, the spatial
perturbation decaying in the slowest way is just the impulsive
contribution associated with the sub-ensemble of fluctuations that
never experienced an internal transition, which propagates back and
forth at constant speed within the system due to the collisions with
the endpoints and relaxing as a function of time as $e^{-\Lambda(t)}$,
see eq.~(\ref{eqx17})). It follows from this observation that the
relaxation function of a generic thermodynamic variable $f(x)$ for a
CTRW-prepared LW ensemble should obey the long-term scaling
\begin{equation}
R_f(t) \sim e^{-\Lambda(t)}\quad .
\label{eq4_3}
\end{equation}
Eq.~(\ref{eq4_3}) suggests that by modulating the functional form of
the transition rates $\lambda(\tau)$ it is possible to predict from LW
dynamics a great variety of relaxation phenomena observed in physical
phenomenology. Specifically, consider for $\lambda(\tau)$ the model
\begin{equation}
\lambda(\tau)= \frac{a \, \beta}{(1+\tau)^{1-\beta}}
\label{eq4_4}
\end{equation}
with $0 < \beta < 1$ and $a>0$. Since $\beta >0$, $\lim_{\tau
  \rightarrow \infty} \tau \, \lambda(\tau)= \infty$,
{$T(\tau)= a \, \beta \, (1+\tau)^{-1+\beta} \exp[1-a
    \, (1+\tau)^{\beta}]$}, and the associated LW process is
{normal diffusive} by possessing the whole hierarchy of
moments $\langle \tau^n \rangle$. The Central Limit Theorem applies,
and its qualitative propagation along ${\mathbb R}$ is, in the
long-term limit, qualitatively identical to the classical mathematical
Brownian motion, whose overall probability density $P(x,t)$ satisfies
the parabolic diffusion equation.

Since $\Lambda(t)= a \left [(1+t)^\beta -1 \right ]$, eq.~(\ref{eq4_3})
indicates that the relaxational decay of any thermodynamic function
$f(x)$ is of the form
\begin{equation}
R_f(t) \sim e^{-a (1+t)^\beta}\quad ,
\label{eq4_5}
\end{equation}
hence there is a stretched exponential decay. This decay, usually
referred to as the Kohlrausch relaxation, is a common feature observed
in many complex systems \cite{stre1,stre2}. Several interpretations
have been proposed for this anomalous behavior \cite{stre4,stre5}, but
to the best of our knowledge this is the first attempt to connect it
to a LW structure of the underlying fluctuations.

Equation~(\ref{eq4_5}) is also interesting from another thermodynamic
perspective. It shows that even LW fluctuations possessing
{normal} diffusive behavior may display highly
non-trivial properties, deviating from the corresponding predictions
of the associated long-term transport model. In the case of the LW
process defined by eq.~(\ref{eq4_4}), the associated transport model,
i.e., the classical hydrodynamic limit of this model, is just the
parabolic diffusion equation for which the relaxation function of a
generic $f(x)$ should decay exponentially as a function of time,
$R_f(t) \sim e^{-\mu_2 \, t}$, where $\mu_2>0$ is the second
eigenvalue of the Laplacian operator equipped with homogeneous von
Neumann conditions at the boundary. {The failure of the
  classical hydrodynamic limit in predicting finer dynamic properties
  for classes of normal diffusive LWs stems from the fact that the
  hydrodynamic limit captures some properties of the LW dynamics,
  specifically the scaling of the mean square displacement, but not
  the entire complexity involved with a LW, which would be obtained by
  considering the whole moment hierarchy.}

\begin{figure}[h]
\begin{center}
{\includegraphics[height=12.5cm]{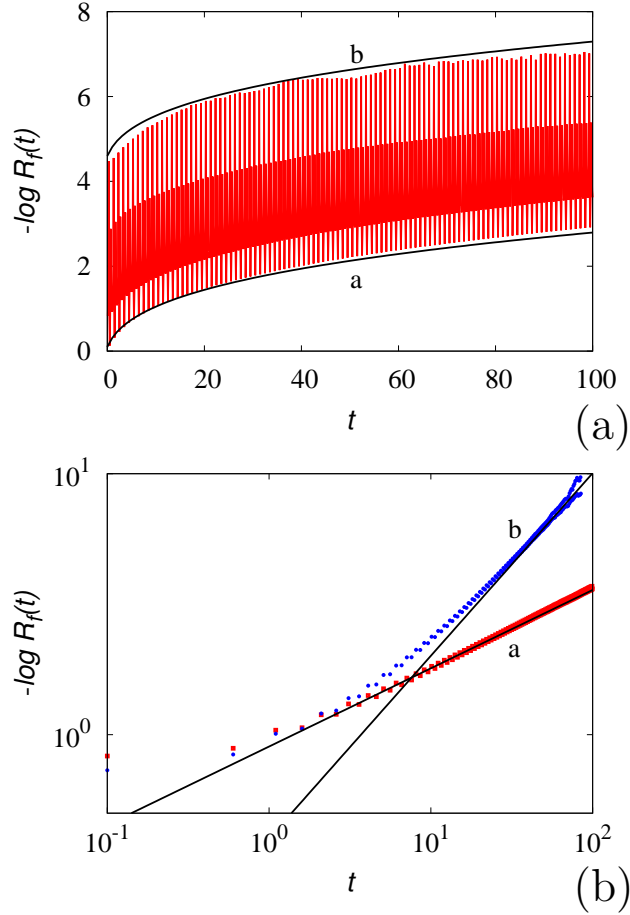}}
\end{center}
\caption{The relaxation function $-\log R_f(t)$ vs.\ $t$ for the
  relaxational dynamics in a closed system discussed in the main text
  using LWs defined by eq.~(\ref{eq4_4}) with $\beta=0.3$, $a=1$.
  Panel (a) shows oscillations in $R_f(t)$ as a function of time as
  explained in the text while the two curves (a) and (b) represent the
  envelope functions $-\log R_f(t)= c (1+t)^\beta$ for two values of
  the prefactor $c$.  Panel (b) depicts the scaling of $-\log R_f(t)$
  once sampled at multiples of the end-to-end transit time $L/b=1$ for
  $\xi=0.3$, $a=1$ (symbols ($\blacksquare$) and curve (a)), and for
  $\xi=0.7$, $a=0.3$ (symbols ($\bullet$) and curve (b)).  Symbols are
  the results of stochastic simulations, lines the scaling curve
  derived from eq.~(\ref{eq4_5}), $-\log R_f(t) \sim (1+t)^\beta +c$,
  where $c$ is a constant.}
\label{Fig4}
\end{figure}

An example of this phenomenon is depicted in Fig.~\ref{Fig4} panel
(a), where the model eq.~(\ref{eq4_4}) is used with $L=1$, $b=1$
[a.u.]. The relaxation data have been derived from stochastic
simulations of the system using $5 \times 10^7$ particles initially
located at $x=x_c=1/2$ with age $\tau=0$ and equiprobable velocity
directions. Statistically, this means that $p_\pm^0(x,\tau) =
\delta(x-x_c) \, \delta(\tau)/2$.  As a thermodynamic test function we
consider a quadratic function of $x$, $f(x)=6 \, x \, (1-x)$, so that
$\overline{f}^*=1$.  Figure~\ref{Fig4} panel (a) shows the time
evolution of the logarithm of $R_f(t)$ with reversed sign for
$\beta=0.3$, $a=1$, displaying the complex oscillations associated
with the back-and-forth propagation of the impulsive mode due to the
reflective boundary conditions and corresponding to the subpopulation
of particles that did not experience any inner transition. The
behavior of $- \log R_f(t)$ is highly nonlinear and bounded from below
and the top by $c_1 \, (1+t)^\beta < R_f(t) < c_2 (1+t)^\beta$, where
$c_1< c_2$ are constant, which is consistent with
eq.~(\ref{eq4_5}). Taking these properties into account, if the
relaxation dynamics is sampled at times $t_n= t_0+ b \, L \, n$,
$n=1,2,\dots$ where $t_0$ is any initial instant of time, a regular
and monotonic behavior in the relaxation dynamics should be observed.
This property is depicted in Fig.~\ref{Fig4} panel (b) for two
different LW systems.

{\subsection{Solute release kinetics}}
\label{sec4_2}

\begin{figure}[h]
\begin{center}
{\includegraphics[height=12.5cm]{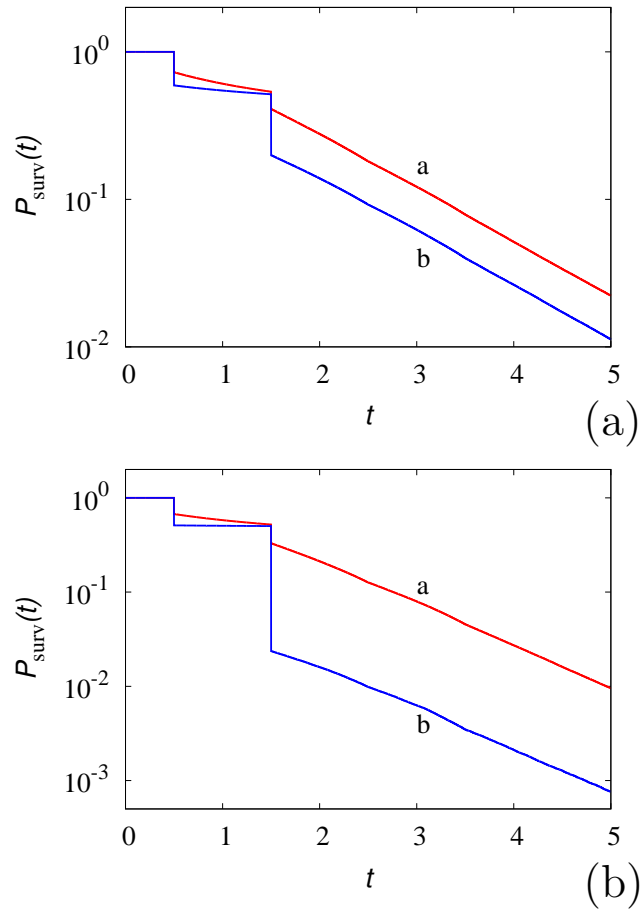}}
\end{center}
\caption {The survival probability density $P_{\rm surv}(t)$ vs.\ $t$ in
  a release experiment in the presence of LW particles whose motion
  is defined by eq.~(\ref{eq2_12}).  Panel (a) refers to
  $\xi=1.5$, panel (b) to $\xi=1.05$.  Lines (a) correspond to the
  equilibrium preparation, lines (b) to the CTRW preparation when
  initially all the particles possess vanishing transitional age.}
\label{Fig5}
\end{figure}

Significant differences controlled by the system preparation occur in
other typical transport experiments involving bounded systems.  Let us
consider the release dynamic of a solute from a complex polymeric
matrix with a transport property that obeys a LW model. Assume that
$x=L$ corresponds to an impermeable boundary to solute transport and
that $x=0$ is the exit boundary from which the solute is released into
the environment. Moreover, assume that the external environment is
perfectly mixed and arbitrarily large so that the solute concentration
outside the release system, and at the exit boundary of it, can be
considered vanishingly small. This transport problem is conceptually
identical to a first passage time problem in which $x=0$ corresponds
to the target exit point \cite{ctrw3,Red01,MOR14}. In the release
experiment, indicating with $P_{\rm surv}(t)$ the fraction of solute
particles still within the release system at time $t$ and with
$J_0(t)$ the particle flux exiting from $x=0$, mass balance dictates
\begin{equation}
P_{\rm surv}(t)= 1- \int_0^t J_0(t^\prime) \, d t^\prime\quad .
\label{eq4_6}
\end{equation}
The flux $J_0(t)$ in the release experiment corresponds exactly to
the first passage time density function $f_{\theta_1}(\theta_1)$
when $t=\theta_1$, i.e.,
\begin{equation}
f_{\theta_1}(\theta_1)= - \left . \frac{d P_{\rm surv}(t)}{d t}\quad 
\right |_{t=\theta_1} . 
\label{eq4_7}
\end{equation}
Consider the LW defined by eq.~(\ref{eq2_12}) with $b=1$ and $L=1$.
Figure \ref{Fig5} depicts the behavior of $P_{\rm surv}(t)$ vs.\ $t$
at short and intermediate time scales for the two values of $\xi=1.05,
\, 1.5$ corresponding to anomalous but transitionally ergodic LW
fluctuations, and for two initial preparations of the system in which
the solute (i.e., the LW particles) are localized initially at
$x=x_c=1/2$ with equiprobable directions of motions and age
distributions corresponding either to the CTRW preparation, i.e.,
$p_{\pm}(x,0,\tau)=\delta(x-x_c)\delta(\tau)/2$, or to the equilibrium
age distribution, i.e., $p_{\pm}(x,0,\tau)=
\delta(x-x_c)e^{-\Lambda(\tau)}/2$.  These data have been obtained
from stochastic simulations starting from an initial ensemble of
$10^8$ solute particles.

One can see that the age preparation of the system deeply modifies the
release properties: The difference in $P_{\rm surv}(t)$ can be of
about two orders of magnitude at $\xi=1.05$ for the two preparations
at time scales when a significant portion of solute is still present
within the system starting from the equilibrium preparation (curve (a)
in Fig.~\ref{Fig5} panel (b)).  Note also the step structure in the
decay of all curves, which is due to the two propagating fronts of the
particle densities and their interplay with the reflecting wall at
$L$, akin to the oscillatory dynamics shown in Fig.~\ref{Fig4}.  In
detail, the first step corresponds to the initial solution propagating
directly towards the exit point $x=0$. Conversely, the second step is
generated by the initial solution propagating in the opposite
direction, which is first reflected at the boundary $x=L$ and only
later reaches the exit point. {Recombination dynamics amongst
the two partial probability waves prevents the occurrence of further jumps for
longer times, and thus a smooth decaying profile sets up.}   While these effects, controlled by the
initial conditions in age, are quantitatively relevant for transport
problems, in the next section we address the peculiarity of the first
passage time problem in the presence of LW fluctuations in the light
of another internal parameter, namely the initial velocity direction,
which plays a leading role as it emerges from the hyperbolic
modelling.

\section{Integral formulation of the first passage time statistics of L\'evy Walks}
\label{sec5}

Let us finally analyze the first passage time problem in the light of
the hyperbolic formulation of the statistical properties of LWs.
Owing to the analogy between LWs and Poisson-Kac processes, it is
convenient to tackle this problem starting from the latter.  While the
main step characterizing the formulation of the hyperbolic transport
equations and of the associated boundary conditions is analogous in
the two cases, LWs may display some anomalies in the long-tail decay
of the first passage time statistics with an exponent differing from
the Sparre-Andersen value of $-3/2$ \cite{DGBD17}, which cannot occur
in the classical Poisson-Kac case defined by eq.~(\ref{eq2_7}).

\subsection{Poisson-Kac processes}
\label{se5_1}

Let us therefore first consider an ensemble of LW particles with
$\lambda(\tau)=\lambda=\mbox{constant}$, initially localized at
$x=x_0>0$ and evolving on the positive semiaxis, and let $x=0$ be the
position of the target exit point.  Once a particle passes through
$x=0$ it is removed from the system, and its first passage time is
evaluated. If $x(t)$ is the continuous trajectory of the particle, the
first passage time $t^*$ is defined as the first time instant for
which $x(t^*-\varepsilon) x(t^*+\varepsilon)<0$ for any small
$\varepsilon>0$, provided that $x(t)>0$ for $t<t^*$. This case,
corresponding to a Poisson-Kac ensemble, is statistically described by
the hyperbolic system of equations for the partial densities
$\widehat{P}_\pm(x,t)$ defined for $x \in (0,\infty)$ and $t>0$.  At
infinity regularity conditions apply, namely $\lim_{x \rightarrow
  \infty} x^k \, \widehat{P}_\pm(x,t)=0$ for any $t>0$ and any
$k=0,1,\dots$. Regarding the condition at the exit point, the above
definition of ``first passage'' implies the removal of any particle
that passes through $x=0$ at any time $t$ from subsequent analysis.
This process involves exclusively $\widehat{P}_+(x,t)$ at $x=0$, which
should necessarily be vanishing, i.e.,
\begin{equation}
\widehat{P}_+(0,t) =0 \quad .
\label{eq5_1}
\end{equation}
Conversely, $\widehat{P}_-(x,t)$ can attain in principle any
non-negative value at $x=0$. The exiting flux at $x=0$ is just
$J_0(t)= b \, \widehat{P}_-(0,t)$ and, consequently, the first
passage time density function $f_{\theta_1}(\theta_1)$ is readily
obtained from the solution of eqs. (\ref{eq2_7}) as
\begin{equation}
f_{\theta_1}(\theta_1)= b \, \widehat{P}_-(0,t) |_{t=\theta_1}\quad .
\label{eq5_2}
\end{equation}
Given the initial conditions
\begin{equation}
\widehat{P}_\pm(x,0)= \pi_\pm^0 \, \delta(x-x_0)\quad ,
\label{eq5_3}
\end{equation}
$\pi_\pm \geq 0$, $\pi_+^0+\pi_-^0=1$, the problem expressed by
eqs.~(\ref{eq2_7}) and equipped with the boundary condition
(\ref{eq5_1}), the regularity condition at infinity, and the initial
condition (\ref{eq5_3}) can be solved easily using Laplace
transforms. The analytic expression for the first passage time
statistics for PK processes has been recently discussed by Rossetto
\cite{rossetto} starting from the Siegert formula \cite{siegert}.  In
order to mark explicitly the dependence on the initial position $x_0$,
it is convenient to indicate the first passage time density as
$f_{\theta_1}(\theta_1;x_0)$.  Although the article by Rossetto
displays some typos in some basic equations, the results are correct
and consequently the analysis is not repeated here.  What is of
relevance in the present analysis are some qualitative observations on
the nature of the first passage time problem of Poisson-Kac
processes. Specifically:
\begin{itemize}
\item This problem is intrinsically vector-valued, in the meaning that
  two first passage time probability densities
  $f_{\theta_1}^{(\pm)}(\theta_1;x_0)$ should be defined accounting
  for the initial preparation of the system with respect to the initial
  velocity orientation: $f_{\theta_1}^{(+)}(\theta_1;x_0)$ is the
  solution of the problem (\ref{eq5_2}) for $\pi_+^0=1$, and
  $f_{\theta_1}^{(-)}(\theta_1;x_0)$ for $\pi_+^0=0$. Owing to
  linearity, the solution for generic initial conditions (\ref{eq5_3})
  is simply
\begin{equation}
f_{\theta_1}(\theta_1;x_0)= \pi_+^0 \, f_{\theta_1}^{(+)}(\theta_1;x_0)+ \pi_-^0
\,  f_{\theta_1}^{(-)}(\theta_1;x_0)\quad .
\label{eq5_4}
\end{equation}
The density $f^{(-)}_{\theta_1}(\theta_1;x_0)$ admits the Laplace transform
\begin{equation}
{\mathcal L}[f^{(-)}_{\theta_1},s]= \exp[-(x_0/b) \sqrt{s} \,
  \sqrt{s+\lambda})]\quad ,
\end{equation}
whose inverse transform is given by
\begin{eqnarray}
f_{\theta_1}^{(-)}(\theta_1;x_0) & = &
e^{-x_0 \, \lambda/b} \, \delta(t-x_0/b)  \nonumber \\
& + & \frac{x_0 \, \lambda}{b} \, e^{-\lambda \, t}
\, \frac{I_1 \left ( \lambda \, \sqrt{t^2 - (x_0/b)^2} \right )}
{\sqrt{t^2 - (x_0/b)^2}} \, \eta(t-x_0/b)\quad ,
\label{eq5_5}
\end{eqnarray}
where $I_1(\xi)$ is the modified Bessel function of the first kind 
of order 1 with argument $\xi$ and $\eta(\xi)$ the Heaviside step
function, $\eta(y)=1$ for $y>0$, $\eta(y)=0$ for $y<0$.
\item Even if initially the particles are located at $x_0=0$,
  i.e., just at the exit point, its first passage time density is not
  necessarily a Dirac delta $\delta(\theta_1)$ provided that $\pi_+^0
  \neq 0$.  Specifically it can be shown that the first passage time
  distribution $f_{\theta_1}^{(+)}(\theta_1;0)$ is given by
\begin{equation}
f_{\theta_1}^{(+)}(\theta_1;0) = \lambda \, e^{-\lambda t} \, \frac{I_1(\lambda \, t)}{\lambda \, t}
\label{eq5_6}
\end{equation}
for which $f_{\theta_1}^{(+)}(\theta_1;x_0)$ follows as
\begin{equation}
f_{\theta_1}^{(+)}(\theta_1;x_0)= \int_0^{\theta_1}
 f_{\theta_1}^{(-)}(\theta_1-\tau;x_0) \, f^{(+)}(\tau;0) \, d \tau
\label{eq5_7}
\end{equation}
admitting a straightforward physical interpretation: The first passage
time density from $x_0$ starting from an initial velocity outwardly
oriented with respect the exit point (i.e.,
$f_{\theta_1}^{(+)}(\theta_1;x_0)$) is the convolution of the
probability density of the time needed to reverse the orientation
(i.e., $f^{(+)}(\tau;0)$) times the first passage time density
from $x_0$ starting from inwardly oriented initial velocities
($f_{\theta_1}^{(-)}(\theta_1-\tau;x_0)$).
\end{itemize}
This setting of the first passage time problem characterizes all the
stochastic processes possessing finite propagation velocity including
LWs.

There is another qualitative issue that distinguishes processes
possessing finite propagation velocity from their Wiener-driven
counterparts. This is associated with the possibility of defining the
first passage time problem from an equivalent transport problem over
the real line using the method of images by locating a suitable
initial condition at the image point $-x_0$ of $x_0$ with respect to
the exit point. Indeed, as observed in \cite{images} this method does
not apply {for L\'evy flights and for anomalously diffusive LWs.}  As a matter of fact, the
method of images fails also for Poisson-Kac processes, and its failure
is not related to the eventual diffusional anomaly of the process but
rather to the boundedness of the propagation velocity, which is
reflected in the symmetry properties of the associated Green functions
for the free-space propagation.

To show this, consider the propagation of eq.~(\ref{eq2_7}) over the
real line with an image condition at the image point $-x_0$,
\begin{equation}
\widehat{P}_\pm(x,0) = \pi_\pm^0 \, \delta(x-x_0) + \pi_\pm^{0,\prime} \, \delta(x+x_0)\quad ,
\label{eq5_8}
\end{equation}
where $\pi_\pm^{0,\prime}$ are unknown real values to be determined by
enforcing the boundary condition (\ref{eq5_1}).  The solution of this
problem can be obtained by using the closed-form expression for the
matrix-valued Green function reported in \cite{giona_pucci}.
Indicating with $(G_{\alpha,\beta}(x,t))_{\alpha,\beta=\pm}$ the
entries of the Green function matrix for an initial condition centered
at $x=0$, the formal solution of the image method reads
\begin{eqnarray}
\widehat{P}_\pm(x,t)&=&G_{+,+}(x-x_0,t) \, \pi_+^0 + G_{+,-}(x-x_0,t)
\, \pi_-^0 \nonumber \\ 
&+ & G_{+,+}(x+x_0,t) \, \pi_+^{0,\prime} + G_{+,-}(x+x_0,t) \, \pi_-^{0,\prime}\quad .
\label{eq5_9}
\end{eqnarray}
At $x=0$ the forwardly propagating density is
\begin{eqnarray}
\widehat{P}_+(0,t) & = & \left [ G_{+,+}(-x_0,t) \, \pi_+^0 +
G_{+,+}(x_0,t) \, \pi_+^{0,\prime} \right ] \nonumber \\
& + & \left [ G_{+,-}(-x_0,t) \, \pi_+^0 +
G_{+,-}(x_0,t) \, \pi_-^{0,\prime} \right ]\quad .
\label{eq5_10}
\end{eqnarray}
Owing to the directed propagation of the Poisson-Kac density waves,
the entries $G_{+,\pm}(x,t)$ are not symmetric functions of their
spatial argument (as can be checked from their explicit analytical
expression reported in \cite{giona_pucci}). Consequently, one
cannot find constants $\pi_\pm^{0,\prime}$ such that the equations
$\pi_\pm^{0,\prime} = - G_{+,\pm}(-x_0,t) \, \pi_\pm^0/
G_{+,\pm}(x_0,t)$ are identically fulfilled for any $t>0$.

\subsection{L\'evy Walks}
\label{sec5_2}
In the case of LWs, the first passage time problem within the
hyperbolic formulation reduces to the solution of eq.~(\ref{eq2_3})
for the partial densities defined in $x \in (0,\infty)$ and equipped
with the boundary condition for the incoming (entering) density wave
\begin{equation}
\left . p_+(0,t,\tau)  \right |_{x=0}=0
\label{eq52_1}
\end{equation}
identical to the corresponding condition (\ref{eq5_1}) for Poisson-Kac
processes.  Indicating with $P_{\rm surv}(t)$ the fraction of
particles remaining in the domain $[0,\infty)$ at time $t$, its
  derivative returns the probability of the first passage times with
  reverse sign.  Enforcing the transformations
\begin{equation}
p_\pm(x,t;\tau) = e^{-\Lambda(\tau)} \, q_\pm(x,t;\tau)
\label{eq52_2}
\end{equation}
the auxiliary functions $q_\pm(x,t;\tau)$ satisfy a
 conservative hyperbolic scheme
\begin{equation}
\frac{\partial q_\pm(x,t;\tau)}{\partial t}= \mp b \, \frac{\partial
q_{\pm}(x,t;\tau)}{\partial x} - \frac{\partial q_\pm(x,t;\tau)}{\partial
\tau}
\label{eq52_3}
\end{equation}
equipped with the boundary conditions
\begin{equation}
q_\pm(x,t;0) = \int_0^\infty T(\tau) \, q_\mp(x,t;\tau) \, d \tau
\label{eq52_4}
\end{equation}
and with the initial conditions $q_\pm(x,0;\tau)=e^{\Lambda(\tau)} \, p^0_\pm(x,\tau)=q^0_\pm(x,\tau)$.

From eq.~(\ref{eq52_3}) it follows that the functional
dependence of the auxiliary functions on their arguments should
necessarily be of the form
\begin{equation}
q_\pm(x,t;\tau)= \phi_\pm(x\ \mp b \, t,t-\tau)\quad .
\label{eq52_5}
\end{equation}
To $q_-(x,t;\tau)$ the same representation used in Sec.~\ref{sec3}
applies, namely
\begin{equation}
q_-(x,t;\tau)=
\left \{
\begin{array}{lll}
q_-^0(x+b \, t, \tau-t) &  & \tau \geq  t \\
h_-(x+b \, \tau,t-\tau) & & \tau < t\quad .
\end{array}
\right .
\label{eq52_6}
\end{equation}
Conversely, the structure of $q_+(x,t)$ should account for the boundary condition
at $x=0$. This can be achieved by setting
\begin{equation}
q_+(x,t;\tau) =
\left \{
\begin{array}{lll}
q_+^0(x-b \, t,\tau-t) & & \tau \geq t \, , \;\, x < b \, t \\
h_+(x-b\, \tau,t-\tau) & & \tau < t\, , \;\, \tau < x/b \\
0 & & \mbox{otherwise}\quad .
\end{array}
\right .
\label{eq52_7}
\end{equation}
The latter representation ensures that no particle that left the
positive region ($x>0$) will re-enter it, which is the fundamental
constraint in order to define correctly the first passage time
statistics.

Substituting these expressions into the boundary conditions
(\ref{eq52_4}), the integral equations for the auxiliary functions
$h_\pm(x,t)$ follow. For $h_+(x,t)$ one derives
\begin{eqnarray}
h_+(x,t)&=& \int_0^t T(t-\tau) \, h_-(x+b \, t - b \, \tau,\tau) \, d \tau \nonumber \\
&+&  \int_t^\infty T(\tau)  \, e^{\Lambda(\tau-t)} \, p_-^0(x+b \, \tau,\tau-t) \, d \tau \quad .
\label{eq52_8}
\end{eqnarray}
For $h_-(x,t)$ eq.~(\ref{eq52_7}) provides
\begin{eqnarray}
h_-(x,t) & = & \int_0^t T(\tau) \, h_+(x-b \, \tau, t-\tau) \, \eta(x- b \, \tau)
\, d \tau  \nonumber \\
& + & \, \eta(x-b \, t) \int_t^\infty  T(\tau) \, e^{\Lambda(\tau-t)}
\, p_+^0(x-b \, t,\tau-t) \, d \tau\quad ,
\label{eq52_9}
\end{eqnarray}
{where $\eta(\cdot)$ is the Heaviside step function.}
Alternatively, the first integral on the r.h.s of eq.~(\ref{eq52_9})
can be expressed as
\[
\int_0^{\mbox{min}\{t,x/b\}} T(\tau) \, h_+(x-b\, \tau,t-\tau) \, d \tau\quad .
\]
The quantity $1-P_{\rm surv}(t)$ represents the distribution function
for the first passage times and the corresponding density follows from
differentiation, see eq.~(\ref{eq4_7}). {$P_{\rm
    surv}(t)$ can be readily obtained from the solution of the above
  integral equations, since by definition $P_{\rm surv}(t)=
  \int_0^\infty d x \int_0^\infty \left [p_+(x,t;\tau)+p_-(x,t;\tau)
    \right ] \, d \tau$.}

\begin{figure}[h]
\begin{center}
{\includegraphics[height=6.5cm]{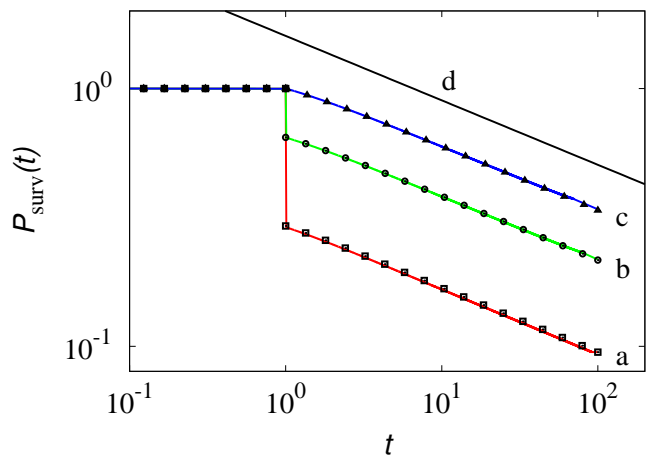}}
\end{center}
\caption{The survival probability $P_{\rm surv}(t)$ vs.\ $t$ for the
  LW defined by eq.~(\ref{eq2_12}), $\xi=0.5$, with $b=1$, $x_0=1$,
  for different values of $\pi_+^0$.  Solid lines (a)-(c) are the
  results of the numerical integration of
  eqs.~(\ref{eq52_8})-(\ref{eq52_9}) with $\Delta t=10^{-2}$, symbols
  the results of stochastic simulations.  Line (a) and ($\square$)
  refers to $\pi_+^0=0$, line (b) and ($\circ$) to $\pi_+^0=0.5$, line
  (c) and ($\triangle$) to $\pi_+^0=1$. The solid line (d) represents
  the scaling $P_{\rm surv}(t) \sim t^{-\xi/2}$.}
\label{Fig6}
\end{figure}

The system of eqs.~(\ref{eq52_8})-(\ref{eq52_9}) has been solved
numerically for a LW with $b=1$ [a.u.] with the transition rate
function expressed by eq.~(\ref{eq2_12}) using $\Delta x= \Delta
t=\Delta \tau=10^{-2}$.  As an initial preparation, consider the case
$p_\pm(x,\tau)= \pi_\pm^0 \, \delta(x-x_0) \, \delta(\tau)$, with
$x_0=1$, and different settings of the probabilistic weights
$\pi_\pm^0>0$, $\pi_+^0+\pi_-^0=1$, controlling the distribution of
the initial velocity directions.  Figure~\ref{Fig6} depicts the
behaviour of $P_{\rm surv}(t)$ vs.\ $t$ obtained from the numerical
solution of the integral Volterra equations for $\xi=0.5$ and
different initial velocity direction distributions, compared with the
corresponding data obtained from the stochastic simulations of the
first passage time problem using an ensemble of $10^8$ particles.
Scaling theory provides for the first passage time density
$f_{\theta_1}(\theta_1) \sim \theta_1^{-\zeta}$, $\theta_1 \gg x_0/b$.
For $\xi>1$ we obtain the exponent $\zeta=3/2$ corresponding to the
Sparre-Andersen result while for $\xi<1$ we get $\zeta=1+\xi/2$
\cite{KoBa11,ACEK14}. In terms of the survival fraction $P_{\rm
  surv}(t)$ this means
\begin{equation}
P_{\rm surv}(t) \sim
\left \{
\begin{array}{lll}
t^{-\xi/2} & & 0 < \xi < 1\\
t^{-1/2} & & \xi > 1\quad .
\end{array}
\right .
\label{eq52_10}
\end{equation}
The integration of the system of equations for $h_\pm(x,t)$ closely
matches the stochastic data and correctly predicts the anomalous
Sparre-Andersen exponent.

\begin{figure}[h]
\begin{center}
{\includegraphics[height=6.5cm]{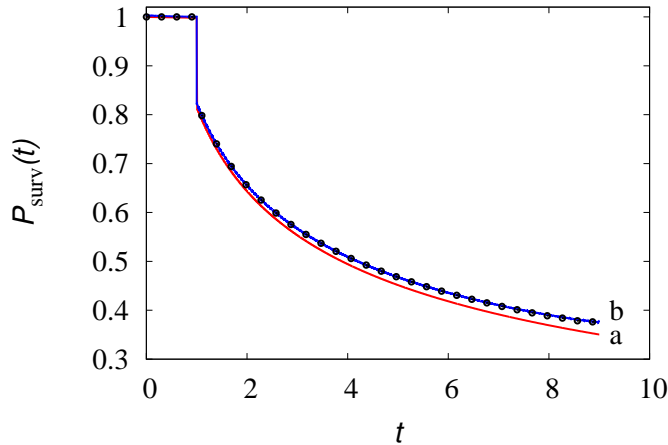}}
\end{center}
\caption{$P_{\rm surv}(t)$ vs.\ $t$ for the LW defined by
  eq.~(\ref{eq2_12}), $\xi=1.5$, with with $b=1$, $x_0=1$ and
  $\pi_0^+=0.5$.  Solid lines represent the result of the numerical
  integration of eqs.~(\ref{eq52_8})-(\ref{eq52_9}) at two different
  step sizes: (a) $\Delta t=10^{-2}$, (b) $\Delta t=10^{-3}$. Symbols
  are the results of stochastic simulations.}
\label{Fig7}
\end{figure}

However, one should be cautious with the numerical integration of
eqs.~(\ref{eq52_8})-(\ref{eq52_9}), as the accuracy may depend
significantly on the step size chosen. This phenomenon is depicted in
Fig.~\ref{Fig7} at $\xi=1.5$ in which a step size of at least $\Delta
t=10^{-3}$ is required for an acceptable prediction of the stochastic
simulation data. This opens up the interesting problem of defining
novel numerical algorithms for the efficient integration of the
integral Volterra equations arising from LW theory, a problem that is
shared by any model expressed in terms of the fractional derivatives
of the overall distribution function $P(x,t)$.

\section{Concluding remarks}
\label{sec6}

The hyperbolic formulation of LWs, parametrized with respect to the
velocity direction and the transitional age, permits to completely
describe their statistical properties in a simple formal setting,
which makes it possible to address a variety of different
phenomenologies within a unified framework. The concept of ensemble
preparation is a direct consequence of this formulation accounting for
the more general case of initial conditions involving the internal
degrees of freedom characterizing LWs. In this framework, the concept
of aging emerges as a particular system preparation.

However, to conceptually simple theoretical settings do not
necessarily correspond computationally simple ways of determining the
respective system properties.  Nevertheless, in the case of the
hyperbolic formulation of the statistical properties of LWs, the extra
degree of freedom represented by the transitional age $\tau$, which
comes in addition to the two variables of space coordinate $x$ and
time $t$ in the partial densities $p_\pm(x,t;\tau)$, can be embedded
within the temporal parametrization. This means that the evolution of
the system can be completely described by means of two auxiliary
function $h_\pm(x,t)$ depending exclusively on a space $x$ and a
temporal coordinate $t$, satisfying a system of Volterra integral
equations in which the convolutional nature of the dynamics accounts
for the memory effects associated with the age. In general, as for
Poisson-Kac processes the parametrization with respect to the velocity
direction, which corresponds to the inclusion of a system of two
partial densities $p_\pm(x,t;\tau)$ (or two auxiliary functions
$h_\pm(x,t)$) in the statistical analysis cannot be eliminated if the
most general initial preparations are considered in which unbalanced
subpopulations of particles initially moving in the two opposite
directions may occur.

If symmetric conditions for the initial population of the forward and
backward moving particles are assumed, only a single auxiliary
function, say $h(x,t)=h_+(x,t)$ is needed in the free-space
propagation, as for any model involving the overall density function
$P(x,t)$, allowing an arbitrary initial age preparation of the
system. The equation for $h(x,t)$ becomes nonlocal, and the
nonlocality reflects the reduction of the model to a single
propagating field. This presents some similarities with the problem of
nonlocality and hidden variables in quantum mechanics, where the
notion of hidden variables is not solely related to the existence of a
``hidden probabilitistic structure'' \cite{bell} but eventually to the
inclusion of neglected propagating fields \cite{raskovski}.

The importance of correctly accounting for the initial preparation of
the system emerges clearly either in the short/intermediate term
dynamics or in problems defined in bounded domains.  Even
diffusionally regular LWs displaying a linear scaling in the mean
square displacement may show interesting anomalous relaxation
properties, such as the occurrence of a stretched exponential
decay. This example suggests a broader application of LW fluctuations
in material science and polymer physics as a model of complex
fluctuations (viscoelasticity, nonlinear viscoelasticity, etc.).

The analysis of the first passage time statistics embedded within the
hyperbolic formalism opens up several interesting pathways for further
investigation. The difficulty with formulating a correct method of
images for the first passage time problem involving LWs is not related
to their anomalous behavior but is intrinsically rooted in the finite
propagation velocity of these fluctuation models.  The mathematical
setting of this problem in terms of partial densities requires that
the partial density associated with an incoming wave from the
surrounding environment should vanish at the target exit boundary in
order not to reinject particles into the domain that have already been
passed through it. The same problem arises in other classes of
dynamics possessing finite propagation velocity such as Poisson-Kac
processes.  The manipulation of the partial density equations,
enforcing the method of characteristics, permits to reduce the
transport problem to the evaluation of two auxiliary functions
$h_\pm(x,t)$ of the spatial coordinate $x$ and the temporal one $t$,
as in the case of the free-space propagation.

The hyperbolic formulation of LWs is particularly suited for modelling
more complex situations, which account for the occurrence of
interparticle interactions, exclusion effects, etc., that in a
mean-field modeling can be described by allowing the velocity $b$ and
the transition rate $\lambda(\tau)$ to depend on the partial wave
densities. This extension has been initiated in
\cite{fed_nonl1,fed_nonl2} for LWs and in \cite{giona_interacting} for
Poisson-Kac processes.\\

{\bf Acknowledgements:} R.K.\ thanks Professors Klapp and Stark (TU
Berlin) for hospitality. He is also grateful for funding from the
Office of Naval Research Global and from the London Mathematical
Laboratory, where he is an External Fellow.  A.C.\ gratefully
acknowledges funding under the Science Research Fellowship granted by
the Royal Commission for the Exhibition of 1851. M.G., A.C.\ and
R.K.\ thank the London Mathematical Laboratory for hospitality during
and after a workshop, where this article was completed.

%\bibliographystyle{plain}
%\bibliography{summ28}

\end{document}